\documentclass[ reprint, 
 aps,
 pra,
 showkeys,
 nofootinbib
]{revtex4-1}
\usepackage{amsfonts}
\usepackage{amssymb}
\usepackage{amsmath}
\usepackage{graphicx}
\usepackage{subfigure}
\usepackage{dcolumn}
\usepackage{bm}
\usepackage{setspace}
\usepackage[hang]{footmisc}
\usepackage[sort&compress]{natbib}
\usepackage{multirow}
\usepackage[sort&compress]{cleveref}
\usepackage{nomencl}
\usepackage[english]{babel}
\usepackage{xcolor, soul}
\usepackage{marginnote}
\usepackage{comment}
 
\sethlcolor{yellow}

\crefname{equation}{\unskip}{\unskip}
\crefname{figure}{\unskip}{\unskip}
\crefname{section}{\unskip}{\unskip}
\crefname{subsection}{\unskip}{\unskip}
\begin{document}
\let\ref\cref

\title{Cylinder-flat contact mechanics with surface roughness}

\author{A. Tiwari}
\affiliation{PGI-1, FZ J\"ulich, Germany}
\affiliation{www.MultiscaleConsulting.com, Wolfshovener str. 2, 52428 J\"ulich}
\author{B.N.J. Persson}
\affiliation{PGI-1, FZ J\"ulich, Germany}
\affiliation{www.MultiscaleConsulting.com, Wolfshovener str. 2, 52428 J\"ulich}

\begin{abstract}
We study the nominal (ensemble averaged) contact pressure $p(x)$ acting on a cylinder squeezed in 
contact with an elastic half space with random surface roughness. 
The contact pressure is Hertzian-like
for $\alpha < 0.01$ and Gaussian-like for $\alpha > 10$, where the
dimension-less parameter $\alpha = h_{\rm rms}/\delta$
is the ratio between the root-mean-square roughness amplitude and the penetration
for the smooth surfaces case (Hertz contact). 
\end{abstract}

\maketitle
\makenomenclature


{\bf 1 Introduction}

The pressure or stress acting in point contacts, e.g., when an elastic ball is squeezed against a nominal flat surface,
or in line contacts, e.g., when an elastic cylinder is squeezed against a flat surface, have many important 
applications, such as the contact of a railway wheel and the rail (point contact), or in O-ring seals (line contact).
In many of these applications, the surface roughness has a big influence on the nominal contact pressure profile.
In a recent study for metallic (steel) seals we found that the maximum of the nominal contact pressure was reduced by
a factor of $\approx 3$ when the surface roughness was taken into account in the analysis\cite{Fisher}. This has a huge influence on the
fluid leakrate and led us to perform a more general study, which we report here, 
of the influence of the surface roughness on the pressure profile for line contacts.

In a classical study Greenwood an Tripp\cite{Green1} (see also Ref. \cite{Green2})
studied the influence of surface roughness on the elastic contact of
rough spheres. They used the Greenwood-Williamson\cite{Green3} (GW) contact mechanics theory where the elastic coupling between the
asperity contact regions is neglected. However, later studies have shown that this coupling is very important
even for small nominal contact pressures, where the distance between the macroasperity contact 
regions may be large\cite{Carb1}.
The reason is that there are smaller asperities (microasperities) on top of the big asperities, and since the contact pressure in the
macroasperity contact region in general is very high, the microasperity contact regions are closely spaced and the elastic coupling between them
cannot be neglected. In the present study we will use the Persson contact mechanics theory which includes the elastic coupling
between all asperity contact regions in an approximate but accurate way\cite{BP,Ref7}.

We note that the GW model is approximately valid if roughness occurs on just one length scale\cite{SSR}. 
Now, when the applied squeezing force is small, the average surface separation, which determines the 
influence of the surface roughness on the nominal contact pressure,
depends mainly on the most long-wavelength roughness component. Thus, for small
applied force the GW theory gives an approximately correct nominal 
pressure distribution if the asperities in the GW model are chosen as the long wavelength roughness part of the roughness 
spectrum. However, this approximation breaks down at high enough
applied force, and cannot describe the area of real contact for any applied force as it depends the whole roughness
spectrum.

Many studies of the contact between rough spheres have been presented\cite{Green1,Green2,India,Dini,Rob,Mus,Plastic}. 
Most of them assume only elastic deformations
but a few studies includes plastic deformations and adhesion. Most recent 
studies are based on numerical methods such as the finite element method,
the boundary element method, or molecular dynamics. However, numerical methods cannot be easily applied to real surfaces of
macroscopic solids, which
typically have roughness extending over very many decades in length scales. Here we will use the Persson
contact mechanics theory\cite{BP,PRL,carb,alm} to study the influence of surface roughness on the 
nominal contact pressure for the contact between a cylinder and a flat (line contact).
For this case we are only aware of one study of the type presented here, 
but using the GW-theory\cite{Cylinder}.

\begin{figure}
\includegraphics[width=0.8\columnwidth]{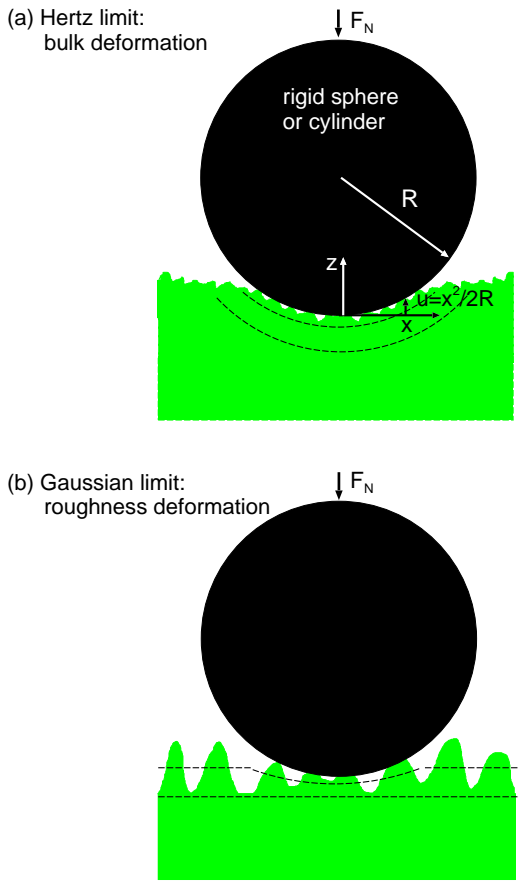}
\caption{\label{TwoLimits1.eps}
Two limiting cases when a rigid cylinder (black) with radius $R$
is squeezed against a nominal flat half-space (green). (a) If the surface
roughness amplitude is very small, or the applied force very high,
the nominal contact area will be determined by bulk deformations and
given by the Hertz contact theory. (b) In the opposite limit mainly the
surface asperities deform (but with a long-range elastic coupling
occurring between them). In this limit the pressure profile is Gaussian-like.
}
\end{figure}

\vskip 0.3cm
{\bf 2 Theory}

Consider an elastic cylinder squeezed against an elastic half-space with the cylinder axis
parallel to the surface of the half-space (flat). The surfaces of the cylinder and the flat have
random roughness with the power spectra $C_1(q)$ and $C_2(q)$, respectively. 
For stationary elastic contact we can map the original problem on another simpler problem, where the cylinder is rigid, 
and the half space elastic with the effective Young's modulus
$${1\over E^*} = {1-\nu_1^2\over E_1} +{1-\nu_2^2\over E_2}.\eqno(1)$$ 
Here $E_1$ and $\nu_1$ are the Young's modulus and Poisson ratio of the cylinder, and  $E_2$ and $\nu_2$
of the half space.
The cylinder surface is perfectly smooth while the flat has
the (combined) surface roughness $h=h_1+h_2$. We assume that the surface roughness on the
two surfaces are uncorrelated so that the combined surface roughness power spectrum
(on the flat) 
$$C(q)=C_1(q)+C_2(q).\eqno(2)$$
Here we have used that $\langle h_1({\bf x}) h_2({\bf x'})\rangle = 0$
(uncorrelated roughness), where $\langle .. \rangle$ stands for ensemble averaging.

From the theory of elasticity\cite{Johns}
$$u=u_1+{x^2\over 2R} -{2\over \pi E^*} \int_{-\infty}^\infty dx' \ p(x')  \Big | {x-x'\over x'} \Big |  \eqno(3)$$
where $p(x)$ and $u(x)$  are the contact pressure, and the interfacial separation, 
averaged over different realizations of the surface roughness. In (3), $u_1$ must be chosen so that 
the applied normal force per unit length $f_{\rm N}$ satisfies
$$\int_{-\infty}^\infty dx \ p(x) = f_{\rm N}\eqno(4)$$
Finally, the Persson contact mechanics theory gives a relation $p(u)$ between the nominal contact pressure
and the interfacial separation $u$, averaged over different realizations of the surface roughness\cite{Yang1}.
For the case of not too high or too low contact pressures, where the system either approach complete contact
or the contact is so small that only a few asperities makes contact (finite-size region)\cite{fs}, we have\cite{PRL}
$$p=\beta E^* e^{-u/u_0},\eqno(5)$$
where $u_0 = \gamma h_{\rm rms}$ (with $\gamma \approx 0.4$) 
and $\beta$ are numbers determined by the surface roughness power spectrum.

If we measure the surface displacement $u$ in units of the sphere or cylinder radius
$R$, and the pressure $p$ in units of the effective modulus $E^*$, we get from (3)-(5):
$$p\approx \beta e^{-u R/u_0},$$
$$u=u_1+{x^2\over 2} -{2\over \pi} \int dx' \ p(x')  \Big | {x-x'\over x'} \Big | , $$
$$\int_{-\infty}^\infty dx \ p(x) = {f_{\rm N} \over RE^*} .$$
Note that the problem depends on the two dimension less parameters
$u_0/R$ and $f_{\rm N}/(RE^*)$. We find it more useful to use instead of $f_{\rm N}/(RE^*)$
the parameter $[u_0/R][RE^*/f_{\rm N}]= \gamma h_{\rm rms}E^*/f_{\rm N}$. Defining 
$\delta = [4/\rm \pi] [f_{\rm N}/E^*]$ the second parameter is essentially $\alpha = h_{\rm rms}/\delta$,
where $\delta$ is the penetration for smooth surfaces (see (9) below).
The parameter $\alpha$ was already introduced  by Greenwood et al\cite{Green1,Green2} in their study of the influence
of roughness on the contact between elastic spheres.

We have solved the equations given above numerically, but two limiting cases can be easily studied
analytically, namely the case of smooth surfaces where the classical Hertz theory is valid, and in the
case where the roughness is very big or the applied force $f_{\rm N}$ small. In the second case we can
neglect the bulk deformations and just include the deformations of the asperities; we refer to this 
limit as the Gaussian limit.

\vskip 0.1cm
{\bf Hertzian limit}

For smooth surfaces ($h_{\rm rms}=0$) the contact is Hertz-like with the pressure distribution\cite{Johns}
$$p=p_0 \left (1-\left ({x\over a}\right )^2 \right )^{1/2}\eqno(6)$$
where
$$p_0 = \left ({E^* f_{\rm N} \over \pi R}\right )^{1/2}\eqno(7)$$
$$a=\left (R\delta \right )^{1/2}\eqno(8)$$
$$f_{\rm N}={\pi \over 4} E^* \delta\eqno(9)$$

For the Hertz contact pressure (6) the ratio between the full width at half 
maximum (FWHM) $w$ and the standard deviation $s$ of $p(x)$ is
easy to calculate: $w/s = 2\surd 3  \approx 3.464$ 

\vskip 0.1cm
{\bf Gaussian limit}

When a cylinder with a smooth surface is squeezed against a flat smooth substrate,
an infinite long rectangular contact region of width $2a$ is formed, with the contact pressure 
given by the Hertz theory (6).
However, if the substrate has surface roughness the nominal contact region will be larger than
predicted by the Hertz theory, and the pressure distribution will change from parabolic-like for the
case of smooth surfaces to Gaussian-like if the surface roughness is large enough. This can be easily
shown using the Persson contact mechanics theory. 

Due to the surface roughness, if the contact pressure
$p$ is not too height the interfacial separation $u$ is related to the contact pressure via (see (5))\cite{Yang1}
$$p=\beta E^* e^{-u/u_0}\eqno(10)$$

Neglecting bulk deformations,
for a cylinder with radius $R$ squeezed against the flat we expect (see (3) with $E^* \rightarrow \infty$): 
$$u \approx u_1+{x^2 \over 2R}\eqno(11)$$
so that
$$p=p_0 e^{-x^2/2s^2}. \eqno(12)$$
where $s^2 = Ru_0 = \gamma  R h_{\rm rms}$. 
Using (12) we get
$$\int_{-\infty}^\infty dx \ p_0 e^{-x^2/2s^2} = p_0 s ( 2\pi )^{1/2} = f_{\rm N}$$
or
$$p_0 = {f_{\rm N} \over s (2\pi )^{1/2}}\eqno(13)$$

We note that (12) holds only as long as the pressure $p$ is so small that the asymptotic relation
(10) is valid, but not too small as then finite size effects become important.
In addition, in deriving (12) we have neglected bulk deformations. This is a valid approximation
only if $s>>a$ or $h_{\rm rms} >> \delta$ or $\alpha = h_{\rm rms}/\delta >> 1$. In this limit
the maximal contact pressure is much smaller than the result for smooth surfaces given by 
the Hertz formula (7). Thus the ratio between (12) and (7) is 
$${p_0 ({\rm rough}) \over p_0({\rm smooth})} = \left ({f_{\rm N} \over 2 \gamma E^* h_{\rm rms}}\right )^{1/2} 
= \left ({\pi \over 8 \gamma \alpha}\right )^{1/2}\eqno(14)$$
Since $\gamma \approx 0.4$ this ratio is about $0.9 \alpha^{-1/2}$.

For a Gaussian function the ratio between the FWHM $w$ and the standard deviation $s$ is
easy to calculate: $w/s = 2 (2 {\rm ln} 2)^{1/2}  \approx 2.355$ 

The study above can also be applied to the contact between a sphere and a flat. In that case
$$u \approx u_1+{r^2 \over 2R}$$
giving
$$p=p_0 e^{-r^2/2s^2}$$
If $F_{\rm N}$ is the applied squeezing force
$$\int d^2x \ p_0 e^{-r^2/2s^2} = F_{\rm N}$$
giving
$$p_0 = {F_{\rm N} \over 2 \pi s^2}.$$

Note that the asperities act like a compliant layer on the surface of the body, so that contact is extended 
over a larger area than it would be if the surfaces were smooth and, in consequence, 
the contact pressure for a given load will be reduced.
In reality the contact area has a ragged edge which makes its measurement subject to uncertainty. 
However, the rather arbitrary definition of the contact width is not a problem when calculating physical quantities like the
leakage of seals, which can be written as an integral involving the nominal pressure (see Ref. \cite{Fisher,liftoff}).

Note also that the fact that the nominal contact pressure $p(x)$ for $\alpha >> 1$ is a Gaussian function of $x$
has nothing to do with the fact that randomly rough surfaces has a Gaussian distribution of asperity heights. Rather,
it results from the fact that there is an exponential relation between the contact pressure and average interfacial separation
(see (5)), and the fact that when bulk deformations can be neglected (which is the case for $\alpha >> 1$) the average interfacial separation
depends quadratically on the lateral coordinate $x$ as long as $x/R <<1$ (see (11)).

\vskip 0.1cm
{\bf Role of plastic deformation}

The derivation of the nominal contact pressure profile presented above assumes elastic deformations. 
If the stress at the onset of plastic flow is higher than the maximum stress $p_0$, and the maximal shear stress, 
 at the surface, and also below the surface
in the cylinder-flat contact region, then no macroscopic plastic deformation will occur. 
In that case, for smooth surfaces we expect no macroscopic plastic deformations, and can treat the contact as
elastic when calculating the nominal contact pressure distribution.
However, the stress in the asperity contact regions is much higher than the nominal contact pressure. 
Thus, assuming elastic contact the relative contact area\cite{Ref7,SSR} $A/A_0 \approx (2/h') (p_0/E^*)$, where $h'$ is the rms-slope. 
Since the average pressure $p$ in the asperity contact regions
must satisfy $p A = p_0 A_0$ we get $p=(A_0/A) p_0\approx h'E^*/2$. 
Thus, at short enough length scale (where $h'$ is large enough) we expect plastic deformations to occur.

However, (12) may still be approximately valid if the asperities deform elastically on the length scale which
determines the contact stiffness for the (nominal) contact pressures relevant for the calculation of (12). The contact stiffness
(or the $p(u)$ relation) for small pressures is determined by the most long-wavelength roughness components
which may deform mainly elastically. Nevertheless, in general
a detailed study is necessary in order to determine the exact influence of plastic flow at the asperity level on the nominal
contact pressure profile.

\begin{figure}
\includegraphics[width=0.95\columnwidth]{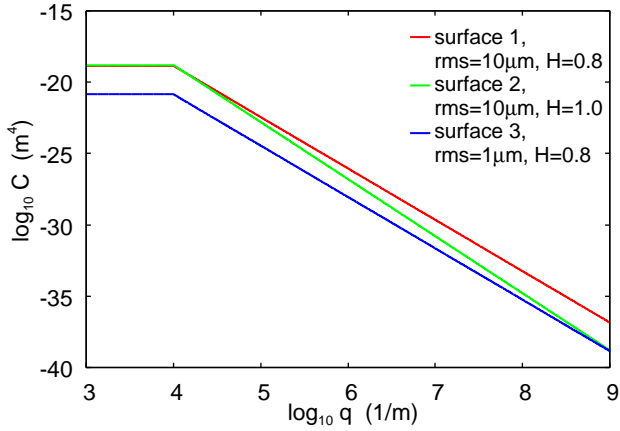}
\caption{\label{1logq.2logC.eps}
The surface roughness power spectra $C(q)$ as a function of the wavenumber $q$ (log-log scale)
for three different surfaces, 1-3, with the root-mean-square (rms) roughness 
amplitude and the Hurst exponents ($h_{\rm rms}=10 \ {\rm \mu m}$, $H=0.8$) (surface 1, red curve)
($10 \ {\rm \mu m}$, $1.0$) (surface 2, green curve) and ($1 \ {\rm \mu m}$, $0.8$) (surface 3, blue curve).
}
\end{figure}

\begin{figure}
\includegraphics[width=0.95\columnwidth]{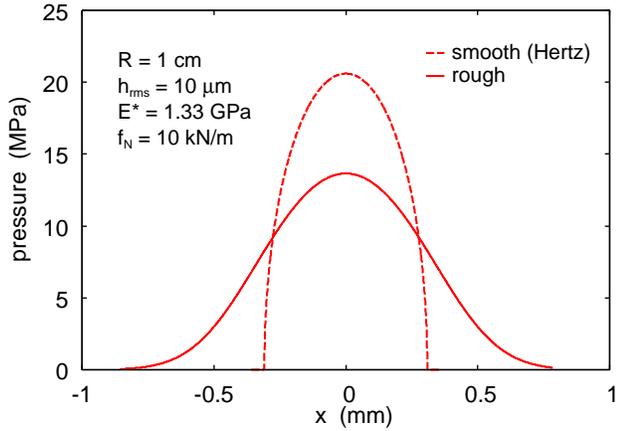}
\caption{\label{1x.2p.Hertz.and.Roughness.eps}
The contact pressure $p(x)$ as a function of the coordinate $x$ for a rigid cylinder
squeezed against an elastic half space. The dashed line is for a
perfectly smooth surfaces (Hertz contact), and the solid line when the substrate
has surface roughness with the power spectrum given by the red line (surface 1) in Fig. \ref{1logq.2logC.eps}. 
The loading force $f_{\rm N} = 10 \ {\rm kN/m}$.
}
\end{figure}

\begin{figure}
\includegraphics[width=0.95\columnwidth]{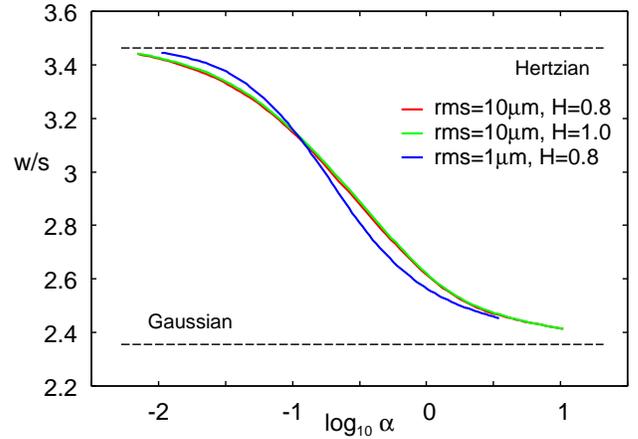}
\caption{\label{1logAlpha.2ratioArea.eps}
The ratio $w/s$ between the full-width-at-half-maximum (FWHM) $w$ and the standard
deviation $s$ as a function of the logarithm of the dimensionless number
$\alpha = h_{\rm rms}/\delta$, where $h_{\rm rms}$ is the rms surface roughness amplitude
and $\delta = 4f_{\rm N}/(\pi E^*)$ the penetration in the Hertz theory (smooth surfaces).
In the calculation we have used the power spectra in Fig. \ref{1logq.2logC.eps}, 
and only varied the loading force $f_{\rm N}$
(between $1000 \ {\rm N/m}$ and $1.5\times 10^6 \ {\rm N/m}$). The cylinder radius $R=1 \ {\rm cm}$
and the effective modulus $E^*=1.33 \ {\rm GPa}$.
}
\end{figure}

\begin{figure}
\includegraphics[width=0.95\columnwidth]{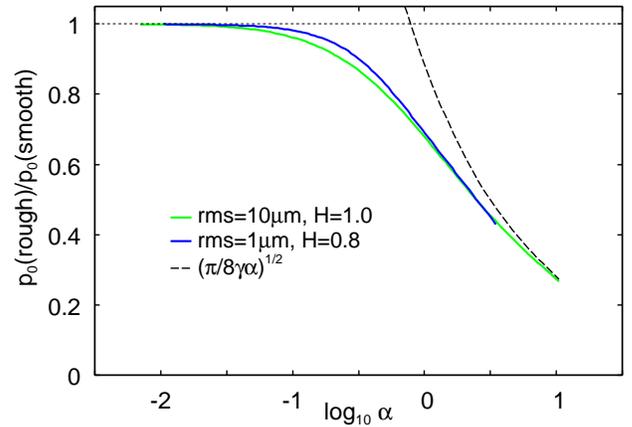}
\caption{\label{1logAlpha.2ratiopmax.eps}
The ratio $p_0({\rm rough})/p_0({\rm smooth})$ between the 
maximum of the contact pressure for rough and smooth surfaces,
as a function of the logarithm of the dimensionless number
$\alpha = h_{\rm rms}/\delta$.
The dashed line is the large $\alpha$ asymtotic result (14).
In the calculation we have used the power spectra in Fig. \ref{1logq.2logC.eps} 
and only varied the loading force $f_{\rm N}$
(between $1000 \ {\rm N/m}$ and $1.5\times 10^6 \ {\rm N/m}$). The cylinder radius $R=1 \ {\rm cm}$
and the effective modulus $E^*=1.33 \ {\rm GPa}$.
}
\end{figure}

\begin{figure}
\includegraphics[width=0.85\columnwidth]{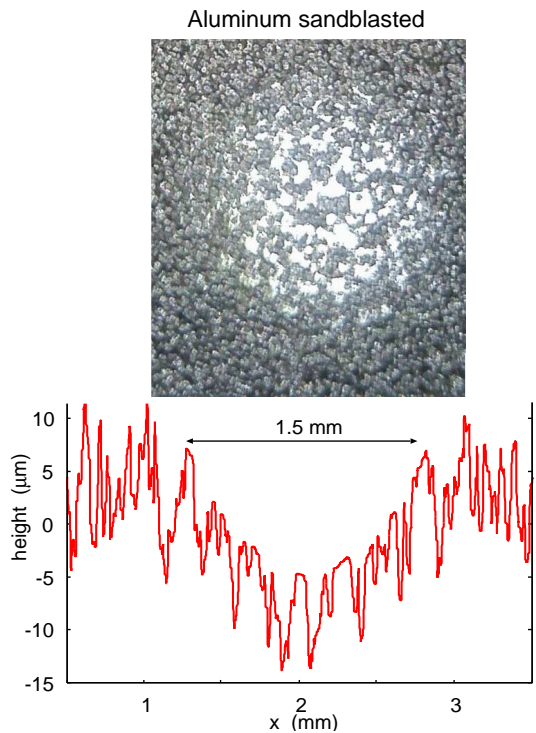}
\caption{\label{AlIndented.2min6bar500NewtonAlumin1.ps}
A sandblasted (6 bar air pressure for 5 minutes) aluminum surface indented by a steel
ball (radius $R=2 \ {\rm cm}$) with $500 \ {\rm N}$ force. Top: The plastically flattened
asperities reflect the light giving the white domains within a circle 
with the diameter $\approx 1.5 \ {\rm mm}$, which is larger than the diameter of the contact
area obtained from the Hertz theory for smooth surfaces (about $0.9 \ {\rm mm}$).
Bottom: stylus line scan through the middle of the indented surface area.
}
\end{figure}

\vskip 0.3cm
{\bf 3 Numerical results and discussion}

We will now present numerical results to illustrate the influence of surface roughness on the
nominal contact pressure profile. In the calculation we will use three surfaces, 1-3, with the roughness power spectra shown in Fig.
\ref{1logq.2logC.eps}. The surface 1 
has the root-mean-square (rms) roughness amplitude $10 \ {\rm \mu m}$, 
and the Hurst exponent $H=0.8$. The other two surfaces have $h_{\rm rms}=10 \ {\rm \mu m}$ and $H=1.0$ (surface 2, green line)
and $h_{\rm rms}=1 \ {\rm \mu m}$ and $H=0.8$ (surface 3, blue line). Note that we include one decade roll-off region, 
from $q=10^3 \ {\rm m}^{-1}$ to $q=10^4 \ {\rm m}^{-1}$. We assume the effective 
Young's modulus $E^*=1.33 \ {\rm GPa}$ and the radius of the cylinder $R=1 \ {\rm cm}$. 
We vary the applied pressure $f_{\rm N}$ from $10^3 \ {\rm N/m}$ to $1.5\times 10^6 \ {\rm N/m}$. This gives 
a variation in $\alpha = h_{\rm rms}/\delta$ with more than 3 decades. 

Fig. \ref{1x.2p.Hertz.and.Roughness.eps} shows the
contact pressure $p(x)$ as a function of the coordinate $x$ for the rigid cylinder
squeezed against an elastic half space with the force $f_{\rm N} = 10 \ {\rm kN/m}$. The dashed line is for 
perfectly smooth surfaces (Hertz contact), and the solid line for the substrate surface 1 with the
surface roughness power spectrum given in Fig. \ref{1logq.2logC.eps} (red line). 
Note that the surface roughness reduces the maximal contact pressure and makes the pressure profile
Gaussian-like, but it becomes a perfect Gaussian only for much larger values of the parameter $\alpha$
(corresponding to smaller applied force $f_{\rm N}$ or larger rms-roughness amplitude $h_{\rm rms}$).

Fig. \ref{1logAlpha.2ratioArea.eps}
shows the ratio $w/s$ between the full-width-at-half-maximum (FWHM) and the standard
deviation $s$ as a function of the logarithm of the dimensionless number
$\alpha = h_{\rm rms}/\delta = (\pi/4) (h_{\rm rms} E^*/ f_{\rm N})$.
In the calculation we have used the three power spectra shown in Fig. \ref{1logq.2logC.eps} 
and varied the loading force $f_{\rm N}$
(between $10^3 \ {\rm N/m}$ and $1.5\times 10^6 \ {\rm N/m}$). 
Note that as a function of $\alpha$ the ratio $w/s$ is nearly the same for all three surfaces.
This suggest that the nominal pressure distribution $p(x)$ depends mainly on the
parameter $\alpha$, as already suggested by Greenwood et al\cite{Green2}.

Fig. \ref{1logAlpha.2ratiopmax.eps}
shows the ratio $p_0({\rm rough})/p_0({\rm smooth})$ between the 
maximum of the contact pressure for the rough and the smooth surfaces,
as a function of the logarithm of the dimensionless number
$\alpha$. We show results for the surfaces 2 and 3 (surface 1 gives nearly the same result as surface 2).
The dashed line is the large $\alpha$ asymtotic result (14), and for $\alpha > 10$ it gives nearly
identical result as obtained for surfaces 1-3.

In this study we have neglected plastic deformation. Plastic deformation may be very important for metals, in particular during
the first contact\cite{plast1,plast2}. As an example, in Fig. \ref{AlIndented.2min6bar500NewtonAlumin1.ps} (top) 
we show the plastically deformed area of a sandblasted aluminum surface
after squeezing (normal force $F_{\rm N} = 500 \ {\rm N}$) 
a steel sphere (radius $R=2 \ {\rm cm}$) with very smooth surface, against the aluminum surface.  Note that asperities have been
smoothed by plastic flow and act as small mirrors resulting in the white regions in the optical picture (observed in reflected light).
In this case some bulk plastic flow has also occured, 
as is clear from the topography line-scan in Fig. \ref{AlIndented.2min6bar500NewtonAlumin1.ps} (bottom),
but similar experiments on a steel surface exhibit only plastic deformations of asperities. 

Note that the contact region is not compact. The diameter of the circular 
region including all the plastically deformed asperities
is $\approx 1.5 \ {\rm mm}$, which is larger than the calculated Hertz contact region 
for smooth surfaces (diameter $\approx 0.9 \ {\rm mm}$).   
In fact, the nominal contact region during the contact with the sphere 
may be even larger than indicated by Fig. \ref{AlIndented.2min6bar500NewtonAlumin1.ps}
because there may be an annular (elastic contact) region outside the plastically deformed region, 
where the contact pressure is too low to induce plastic deformations of the aluminum asperities.
Similar plastic effects may occur in some cases in the line contact problem studied above.

\vskip 0.3cm
{\bf 4 Summary and conclusion}

We have studied the dependency of the nominal contact pressure
on the surface roughness and the loading force when a rigid cylinder is squeezed against
an elastic half space. We found that the contact pressure is Hertzian-like
for $\alpha < 0.01$ and Gaussian-like for $\alpha > 10$, where the
dimension-less parameter $\alpha = h_{\rm rms}/\delta$
is the ratio between the root-mean-square roughness amplitude and the penetration
for the smooth surfaces case (Hertz contact).

\vskip 1.5cm
{\it Acknowledgments:}
This work was funded by the German Research Foundation (DFG) in the scope of the
Project "Modellbildung metallischer Dichtsitze" (MU1225/42-1). The authors would like to thank DFG for its support.


\begin{thebibliography}{99}

\bibitem{Fisher}
F.J. Fischer, K. Schmitz, A. Tiwari and B.N.J. Persson,
{\it Fluid leakage in metallic seals},
subm. to Tribology Letters (2020).

\bibitem{Green1}
J.A. Greenwood and J.H. Tripp,
{\it The elastic contact of rough spheres},
J. Appl. Mech. {\bf 89}, 153 (1967).

\bibitem{Green2}
J.A. Greenwood, K.L. Johnson and E. Matsubara,
{\it A surface roughness parameter in Hertz contact},
Wear {\bf 100}, 47 (1984).

\bibitem{Green3}
J.A. Greenwood, J.B.P Williamson, 
{\it Contact of nominally flat rough surfaces},
Proc R Soc Lond A {\bf 295}, 300 (1966).

\bibitem{Carb1}
G. Carbone, F. Bottiglione,
{\it Asperity contact theories: Do they predict linearity between contact area and load?},
Journal of the Mechanics and Physics of Solids {\bf 56}, 2555 (2008).

\bibitem{BP}
B.N.J. Persson
{\it Theory of rubber friction and contact mechanics},
The Journal of Chemical Physics {\bf 115}, 3840 (2001).

\bibitem{Ref7}
Martin H M\"user, Wolf B Dapp, Romain Bugnicourt, Philippe Sainsot, Nicolas Lesaffre, Ton A Lubrecht, Bo NJ Persson, Kathryn Harris, Alexander Bennett, Kyle Schulze, Sean Rohde, Peter Ifju, W Gregory Sawyer,
Thomas Angelini, Hossein Ashtari Esfahani, Mahmoud Kadkhodaei, Saleh Akbarzadeh, Jiunn-Jong Wu, Georg Vorlaufer, Andras Vernes, Soheil Solhjoo,
Antonis I Vakis, Robert L Jackson, Yang Xu, Jeffrey Streator, Amir Rostami, Daniele Dini,
Simon Medina, Giuseppe Carbone, Francesco Bottiglione, Luciano Afferrante, Joseph Monti, Lars Pastewka, Mark O Robbins, James A Greenwood,
{\it Meeting the contact-mechanics challenge},
Tribology Letters {\bf 65}, 118 (2017).


\bibitem{SSR}
B.N.J. Persson,
{\it Contact mechanics for randomly rough surfaces},
Surface science reports {\bf 61}, 201 (2006).

\bibitem{India}
N. Mulakaluri and B.N.J. Persson,
{\it Adhesion between elastic solids with randomly rough surfaces: Comparison 
	of analytical theory with molecular-dynamics simulations},
Europhysics Letters {\bf 96}, 66003 (2011).


\bibitem{Dini}
 Medina S., Dini D., 
 {\it A numerical model for the deterministic analysis of adhesive rough contacts down to the nanoscale}, 
 Int. J. Solids Struct. {\bf 51}, 2620 (2014).
 
\bibitem{Rob}
 L. Pastewka and M.O. Robbins,
 {\it Contact area of rough spheres: Large scale simulations and simple scaling laws},
 Appl. Phys. Lett. 108, 221601 (2016)
 
\bibitem{Mus}
 M.H. M\"user,
 {\it On the Contact Area of Nominally Flat Hertzian Contacts},
 Tribology Letters {\bf 64}, 11249 (2016).
 
 
\bibitem{Plastic}
 M. Bahrami, M.M. Yovanovich, and J.R. Culham 
 {\it A Compact Model for Spherical Rough Contacts},
 Journal of Tribology,  October (2005).

\bibitem{fs}
L. Pastewka, N. Prodanov, B. Lorenz, M.H. M\"user, M.O. Robbins, B.N.J. Persson,
{\it Finite-size scaling in the interfacial stiffness of rough elastic contacts},
Phys. Rev. E {\bf 87}, 062809 (2013).

 
\bibitem{PRL}
B.N.J. Persson,
{\it Relation between Interfacial Separation and Load: A General Theory of Contact Mechanics},
Phys. Rev. Lett. {\bf 99}, 125502 (2007).

\bibitem{carb}
L. Afferrante, F. Bottiglione, C. Putignano, B.N.J. Persson, G. Carbone,
{\it Elastic contact mechanics of randomly rough surfaces: an assessment of advanced asperity models and Persson's theory},
Tribology Letters {\bf 66}, 75 (2018).

\bibitem{alm}
A. Almqvist, C. Campan, N. Prodanov and B.N.J. Persson,
{\it Interfacial separation between elastic solids with randomly rough
	surfaces: Comparison between theory and numerical techniques},
Journal of the Mechanics and Physics of Solids {\bf 59}, 2355 (2011).
 
\bibitem{Cylinder}
C.C. Lo,
{\it Elastic contact of rough cylinders},
International Journal of Mechanical Sciences {\bf 11}, 105 (1969).

\bibitem{Johns}
K. L. Johnson,
{\it Contact mechanics},
Cambridge university press (1987).

\bibitem{Yang1}
C. Yang and B.N.J. Persson: 
{\it Contact mechanics: contact area and interfacial separation from small contact to full contact}, 
J. Phys.: Condens. Matter {\bf 20}, 215214 (2008).
 
\bibitem{liftoff}
B. Lorenz and B.N.J. Persson,
{\it Time-dependent fluid squeeze-out between solids with rough surfaces},
The European Physical Journal E {\bf 32}, 281 (2010).



\bibitem{plast1}
L. Kogut, I. Etsion,
{\it Elastic-plastic contact analysis of a sphere and a rigid flat},
J. Appl. Mech. {\bf 69}, 657 (2002).

\bibitem{plast2}
R.L. Jackson, I. Green,
{\it A finite element study of elasto-plastic hemispherical contact against a rigid flat},
J. Trib. {\bf 127}, 343 (2005).


%
%
%
%
%
%
%
%
%
%
%
 
\end{thebibliography}
\end{document}